\documentclass[aps,prd,showpacs,nobibnotes,twocolumn,aps,nofootinbib,preprintnumbers,letterpaper]{revtex4}
\usepackage{epsfig}
\usepackage{amsmath}
\usepackage{amssymb}
\usepackage{graphicx}% Include figure files
\usepackage{dcolumn}% Align table columns on decimal point
\usepackage{bm}% bold math

\def\refb#1{(\ref{#1})}
\newcommand{\be}{\begin{equation}}
\newcommand{\ee}{\end{equation}}
\newcommand{\ba}{\begin{eqnarray}}
\newcommand{\ea}{\end{eqnarray}}
\def\refb#1{(\ref{#1})}

\def\laq{\raise 0.4ex\hbox{$<$}\kern -0.8em\lower 0.62ex\hbox{$\sim$}}
\def\gaq{\raise 0.4ex\hbox{$>$}\kern -0.7em\lower 0.62ex\hbox{$\sim$}}
\begin{document}

\preprint{FERMILAB-PUB-07-105-A}

\title{Impact of astrophysical processes on the gamma-ray background from dark
matter annihilations}
\author{Eun-Joo Ahn$^{a,b}$, Gianfranco Bertone$^c$, David Merritt$^d$, Pengjie
Zhang$^{e}$}
%    \altaffiliation{Email: sein@oddjob.uchicago.edu}
\affiliation{$^a$ Department of Astronomy \& Astrophysics 
and Kavli Institute for Cosmological Physics, \\
The University of Chicago, Chicago, IL, USA, \\
NASA/Fermilab Theoretical Astrophysics Group, Batavia, IL, USA}
\affiliation{$^b$ Bartol Research Institute, Department of Physics and
Astronomy, University of Delaware, Newark, DE, USA}
%
%\author{Gianfranco Bertone}
%  \altaffiliation{Email: gianfranco.bertone@pd.infn.it}
\affiliation{$^c$ INFN, Sezione di Padova, Via Marzolo 8, Padova I-35131, Italy}
%
%\author{David Merritt}
%   \altaffiliation{Email: merritt@astro.rit.edu}
\affiliation{$^d$Department of Physics, Rochester Institute of Technology,
Rochester, NY, USA}
%
%\author{Pengjie Zhang}
%    \altaffiliation{Email: pjzhang@shao.ac.cn}
\affiliation{$^e$ Shanghai Astronomical Observatory, Chinese Academy of
Science, Shanghai, China, 200030}

\date{\today}

\begin{abstract}
We study the impact of astrophysical processes on the gamma-ray background
produced by the annihilation of dark matter particles in cosmological halos,
with particular attention to the consequences of the formation of supermassive
black holes. In scenarios where these objects form adiabatically from the
accretion of matter on small seeds, dark matter is first compressed into very
dense ``spikes'', then its density progressively decreases due to annihilations
and scattering off of stellar cusps. With respect to previous analyses, based on
non-evolving halos, the predicted annihilation signal is higher and
significantly distorted at low energies, reflecting the large contribution to
the total flux from unevolved spikes at high redshifts. The peculiar spectral
feature arising from the specific redshift distribution of the signal, would
discriminate the proposed scenario from more conventional astrophysical
explanations. We discuss how this affects the prospects for detection and
demonstrate that the gamma-ray background from DM annihilations might be
detectable even in absence of a signal from the Galactic center. 
\end{abstract}

\pacs{95.35.+d, 97.60.Lf, 98.62.Gq}

\maketitle 

\section{Introduction}

Indirect dark matter (DM) searches are based on the detection of  secondary
particles such as gamma-rays, neutrinos and anti-matter, produced by the
annihilation of DM particles either directly, or through the fragmentation
and/or decay of intermediate particles  (for recent reviews  see
Refs.~\cite{Bergstrom:2000pn, Munoz:2003gx, Bertone:2004pz}).

Among the proposed strategies of indirect detection, searching for a diffuse 
gamma-ray background produced by the annihilation of DM in all halos at all
redshifts appears particularly interesting, because of the useful information
that such a signal would provide on the distribution and evolution of dark
matter halos \cite{Bergstrom:2001jj, Taylor:2002zd, Ullio:2002pj}. Previous
calculations have been performed under the hypothesis that the shape of DM
profiles doesn't change with time, a circumstance that led to the conclusion
that the prospects of detecting gamma-rays from the Galactic center (GC) are
more promising than the gamma-ray background \cite{Ando:2005hr}. However, the
annihilation signal mainly comes from the innermost regions of the DM halos,
i.e. regions where the gravitational potential is dominated by baryons, and
where the extrapolation of numerical simulations is most uncertain. 

In particular, the strong evidence for supermassive black holes (SMBHs) at the
centers of galaxies suggests that the DM  profile is inevitably affected by
astrophysical processes on scales that cannot be resolved by numerical
simulations \cite{Merritt:2006cr}. The formation of massive black holes (BHs) at
the centers of DM halos can significantly modify the DM profile, especially if
the process of BH formation happens ``adiabatically'', i.e. the formation
timescale is much longer than the dynamical timescale of DM particles around it
\cite{peebles:1972, young:1980, Ipser:1987ru, Quinlan:1995, Gondolo:1999ef}.
These so-called {\it spikes} of DM inevitably interact with stars and other
structures in the Universe (e.g. binary black holes), a circumstance that
typically leads to a decrease of the DM  density, and thus of the annihilation
signal \cite{Ullio:2001fb, Merritt:2003qk, Bertone:2005hw, Bertone:2005xv}.

In order to detect the enhancement of annihilation radiation from these dense
structures, one thus has to look either for spikes where astrophysical 
processes are less effective, that evolve in regions with low baryonic
densities, as in the case of {\it intermediate-mass black holes}
\cite{Bertone:2005xz}, or for the contribution to the gamma-ray background from
spikes at high redshift, when the DM enhancements had not yet been depleted by 
astrophysical processes.

It is therefore important to re-analyze the prospects for detecting the
gamma-ray background produced by cosmological DM annihilations, in a
self-consistent scenario that takes into account the time-dependent effect of 
astrophysical processes on the distribution of DM. Here, we first provide a
prescription to assign BH masses and stellar cusps to generic halos of any mass
and at any redshift. We then follow the formation of spikes around SMBHs at high
redshift, and their subsequent disruption due to the interaction with the
stellar cusp and to DM annhilations. Finally, we integrate the annihilation
signal over all redshifts and all structures and discuss the prospects for
detecting the induced gamma-ray background.  

The paper is organized as follows: In Sec.~\ref{sec:smbh} we specify how to
assign spikes to cosmological halos of given mass and a given redshift, and how 
spikes evolve. In Sec.~\ref{sec:bgnd} we calculate the gamma-ray background
produced by DM annihilations in halos of all masses and at all redshifts.
Finally in  Sec.~\ref{sec:conclu} we present our conclusions. We include the
description of the halo density profile, its mass distribution, and evolution
for the sake of completeness and to allow comparison with existing literature in
Sec.~\ref{sec:halo}. Throughout this paper, we assume a flat $\Lambda$CDM
cosmology with $\Omega_m = 0.3$, $h = 0.65$, spectral index $n = 1$, and
$\sigma_8 = 0.9$.

\section{Assigning Spikes to halos}
\label{sec:smbh}

To estimate the effect of BHs on the gamma-ray background produced by DM
annihilations, we first need to model the formation and evolution of BHs in
halos of given mass and at a given redshift, and to follow the formation of  DM
spikes, and their subsequent destruction due to scattering off stars  and to DM
annihilations. Strong constraints on the BH population at all redshifts come
from the relationships between DM halo properties and BH  masses observed in the
local universe, and from the quasar luminosity  function. In this section we
devise a strategy to assign BH masses to host halos at any redshift, and to
calculate the DM distribution in the resulting spikes. Further details on the
normalization of DM halos, and on the  their cosmological evolution, can be
found in the Appendix.

\subsection{SMBH formation}

In $\Lambda$CDM cosmologies, DM halos ($\sim 10^8 \, M_{\odot}$) begin to form
at large redshifts ($z\sim 20$) and subsequently grow through mergers, while
stars form from gas that falls into the halo potential wells. At some point,
SMBHs form from the stars and gas at the centers of the halos. Exactly how this
occurs is not clear. However, the luminosity function of quasars as a function
of redshift traces the accretion history of these BHs \cite{Soltan:1982vf},
suggesting that BHs grew significantly, by accretion, from their initial seeds,
with large mass-to-energy conversion efficiency \cite{Yu:2002sq, Elvis:2001bn,
Marconi:2003tg}. An estimate of the average growth history of BHs presented in
Ref.~\cite{Marconi:2003tg}, suggests that the redshift by which BHs have reached
50\% of their current mass, varies with the BH mass, ranging from $z>2$, for BHs
more massive than $10^{10} \,M_\odot$, to $z<1$ for BH masses below $10^{6}
\,M_\odot$. We adopt here a simplified approach, where {\it all} BHs were
already in place at a characteristic redshift of formation $z=z_{BH}$, and will
discuss the dependence of our results on $z_{BH}$.

In the local universe, tight empirical relations are observed between SMBH mass
and the mass of the DM halo \cite{Ferrarese:2002ct} and the luminosity
\cite{Marconi:2003hj} and velocity dispersion \cite{ff05} of the stellar
component. Based on these results, we adopted the following prescription for
assigning SMBHs to halos:
\begin{enumerate}
\item  The local correlation between SMBH and halo mass \cite{Ferrarese:2002ct}
is used to calculate the mass of the SMBH ($M_{SMBH}$) lying in a halo of mass
$M$ at $z=0$.
\item A SMBH with this mass is placed in the progenitor of this halo at
$z=z_{BH}$.
\item The halo is evolved from redshift $z_{BH}$ to 0 \cite{Wechsler:2001cs},
while leaving $M_{SMBH}$ fixed.
\end{enumerate}

\begin{figure}
\begin{center}
\epsfig{file=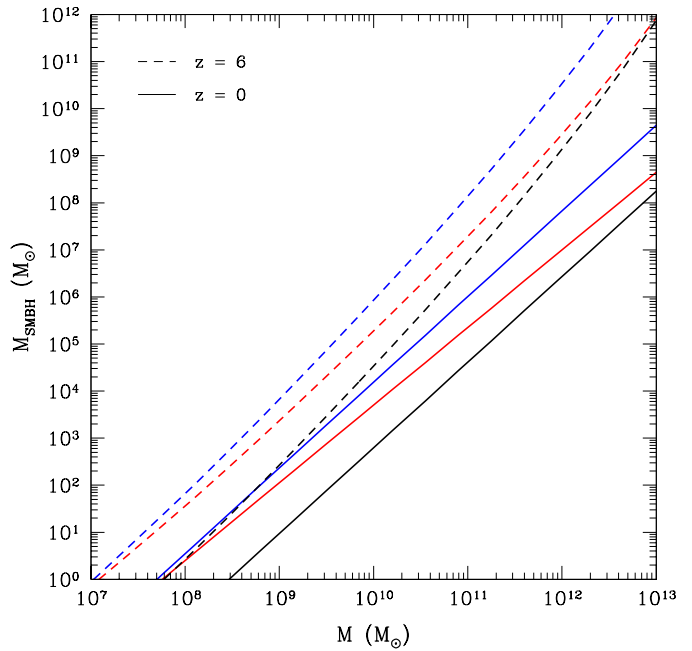,width=0.5\textwidth}
\end{center}
\vskip -2em
\caption{$M_{SMBH}$ as a function of the host halo mass $M$ at $z=6$ (dashed
lines) and $z=0$ (solid lines). The three relations in Eq.~\refb{msmbhm}
$(a)$-$(c)$ are shown from bottom to top for both redshifts.}
\label{figmsmbh}
\end{figure}

Based on Ref.~\cite{Ferrarese:2002ct}, we considered the following relations
between $M_{SMBH}$ and $M$ at $z=0$:
\ba
{ M_{SMBH} \over 10^8 M_\odot } \,=\,
\begin{cases}
~0.027 \, (M_{12,0})^{1.82} &~~~~~(a)\\
~~0.10 \; (M_{12,0})^{1.65} &~~~~~(b)\\
~~0.67 \; (M_{12,0})^{1.82} &~~~~~(c)
\end{cases} \,,
\label{msmbhm}
\ea
where $M_{12,0} \equiv M(z=0) / 10^{12} \, M_\odot$. The differences reflect
different assumptions between the virial radius $r_v$ and the circular velocity.
Figure \ref{figmsmbh} shows the above three relations between $M_{SMBH}$ and
$M$.
The $M_{SMBH}$ obtained at $z=0$ is subsequently placed in halos at 
$z=z_{BH}$. For a fixed halo mass, SMBH at $z=0$ is less massive than SMBH at
$z=6$. This is because the halos at $z=6$ would have evolved to a more massive
halo by $z=0$, where the $M_{SMBH}$ is determined. 

\subsection{Formation and evolution of DM Spikes}
\label{sec:spikes} 

The growth of SMBHs inevitably affects the surrounding distribution of DM. In
fact, it can be shown that the  adiabatic growth of a massive object at the
center of a  power-law distribution of matter with index $\gamma_c$, induces  a
redistribution of matter into a new, steeper, power-law with index $\gamma_{sp}
= 2 + 1/(4-\gamma_c)$ \cite{peebles:1972, young:1980,Ipser:1987ru, Quinlan:1995,
Gondolo:1999ef}. Such a DM enhancement is usually referred to as a ``spike''. 
For the widely adopted Navarro, Frenk and White (NFW) profile (see Appendix for
further comments and references), $\gamma_c=1$, and the spike profile,
immediately after its formation (i.e. at $t=t_f$, the time when the spike is
formed), can be expressed as
\be
\rho_{sp}(r,0) = \rho(r_{b,0}) \left (\frac{r}{r_{b,0}} \right)^{-7/3}
\label{spike}
\ee
inside a region of size $r_{b,0} \approx 0.2 \, r_h$ \cite{Merritt:2005yt}, 
where $r_h$ is the radius of gravitational influence of the SMBH that is 
defined as 
\be
r_h \,\equiv\, {G M_{SMBH} \over \sigma^2} \,,
\label{rh}
\ee
where $G$ is Newton's constant and $\sigma$ the one-dimensional velocity
dispersion. $M_{SMBH}$ can be related to $\sigma$ through the empirical
relation \cite{ff05}
\be
\displaystyle 
{M_{SMBH} \over 10^8 \, M_\odot} \,=\, (1.66 \pm 0.24) \, \left( { \sigma 
\over 200 \, \textrm{km s}^{-1} } \right)^{4.86 \pm 0.43} \,.
\label{msmbhsigma}
\ee
Eq.~\refb{msmbhsigma} is known to be valid for SMBHs in the mass range $10^{6.5}
M_\odot \lesssim M_{SMBH} \lesssim 10^{9.5}M_\odot$ and may extend to higher and
lower masses \cite{ff05}.

Once the spike is formed, several particle physics and astrophysical  effects
tend to destroy it (e.g.~\cite{Ullio:2001fb, Merritt:2003qk, Bertone:2005hw,
Bertone:2005xv}). Here we focus on the gravitational interaction between DM and
stars near the SMBH, which causes a damping of the spike, and on 
self-annihilation of DM near the SMBH, which decreases the maximum  density of
the spike.

The DM and baryons gravitationally interact with each other. Stars in galactic
nuclei have much larger kinetic energies than DM particles, and gravitational
encounters between the two populations tend to drive them toward mutual
equipartition. DM is thus heated up, dampening the spike while maintaing roughly
the same shape of the density profile. Based on the results in
Refs.~\cite{Merritt:2003qk, Bertone:2005hw}, we adopted the following
approximate expression for the decay of the spike intensity with time:
\be
\rho(r,t) \,\approx\, \rho(r,0) \, \kappa  \,;~~~~ \kappa \equiv e^{-\tau/2} \,,
\label{rhoevol1}
\ee
where $\tau$ is the time since spike formation in units of the heating time
$T_{heat}$ \cite{Merritt:2003qk}
\begin{multline}
\displaystyle
T_{heat} \,=\, 1.25 {\rm Gyr} \, \times \\
\left( {M_{SMBH} \over 3 \times 10^6 M_\odot} \right)^{\frac{1}{2}}  \left({r_h
\over 2 {\rm pc} } \right)^{\frac{3}{2}}  \left( {M_\odot \over \tilde{m}_\star}
\right)  \left( {15 \over \ln \Lambda} \right) .
\label{theat}
\end{multline}
$\tilde m_\star$ is the effective stellar mass, and is equal to $\sim 1.8
M_\odot$ assuming a Salpeter mass function and $0.08M_\odot\le m_\star\le
20M_\odot$. $\ln \Lambda = \ln (0.4N)$, with $N \approx 6 \times 10^6$ the
number of stars within $r_h$ for our Milky Way Galaxy. Although $r_h$ is a
function of $M_{SMBH}$, we approximate $\ln \Lambda$ to be constant as
$T_{heat}$ dependence on $N$ is logarithmic.

The size of the spike decreases, with respect to the initial value $r_{b,0}$,
with time
\be
r_b(t) \,=\, \kappa^\delta \, r_{b,0} \,,~~~ \delta \,\equiv\, ( \gamma_{sp} -
\gamma_c)^{-1} \,.
\label{spike-rb}
\ee
The spike density profile is thus given as
\be
\displaystyle
\rho_{sp}(r,t) \,=\, \rho_{sp,0} \, \kappa^\epsilon \, \left({r \over
r_b} \right)^{-\gamma_{sp}} ~,~~~ \epsilon \equiv \gamma_c / (\gamma_c -
\gamma_{sp}) \,.
\label{densitysp}
\ee

\begin{figure}
\begin{center}
\epsfig{file=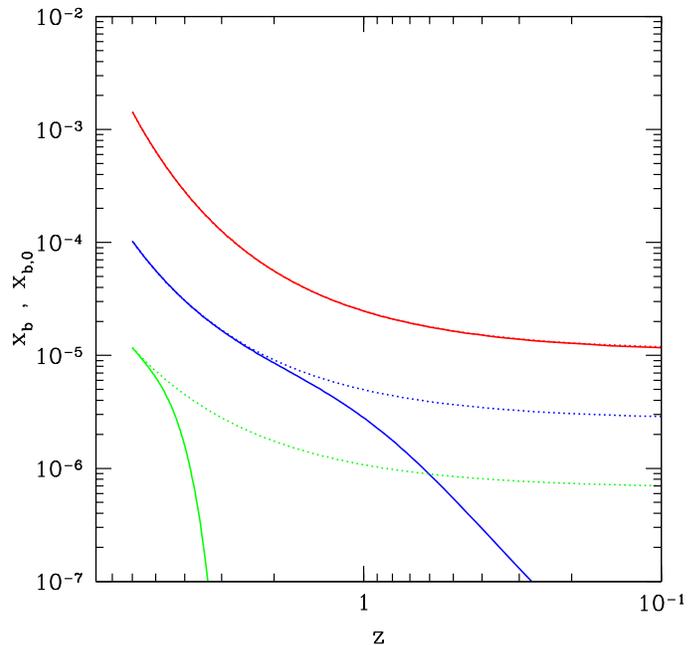,width=0.5\textwidth}
\end{center}
\vskip -2em
\caption{Redshift evolution of the spike parameters $x_b$ (solid lines) and
$x_{b,0}$ (dotted lines), for various halo masses: $M_h(z_{BH}) = 10^{13}, ~
10^{12}, ~ 10^{11} ~M_\odot$ from top to bottom (red, blue, green), and assuming
$z_{BH}=6$.} 
\label{fig:xbxbo}
\end{figure}
Figure \ref{fig:xbxbo} shows the spike evolution under the effect of DM
interaction with baryons, assuming $z_{BH}=6$ for different halo masses. The
spike parameters are shown in units of the halo reference radius $r_0$ (defined
in Sec.~\ref{sec:halo}), so that $x_b \equiv r_b/r_0$ and $x_{b,0} \equiv
r_{b,0}/r_0$. It can be seen that spikes formed in very massive halos are not
affected by heating from the baryons, whereas those in less massive halos
quickly dissipate. Hence low mass halos give negligible contribution to the
gamma-ray signal.

A robust lower limit on the size of the spike is provided by the last stable
orbit ($r_{lso}$) of a test particle around the SMBH. However, annihilation
itself sets an upper limit on the DM density. The evolution equation of DM
particles at radius  $r$ and time $t$ is 
\be
\dot{n}_{sp}(r,t) \,=\, - \langle \sigma v \rangle \, n_{sp}(r,t)^2 \,,
\ee
where the dot denotes a time derivative. Although this expression is  correct
for circular orbits, a more sophisticated approach would take into account the
eccentricities of orbits, and would start from the single-particle distribution
function $f(E,L)$ describing the DM particles, where $E$ and $L$ are the energy
and angular momentum per unit mass respectively, and compute orbit-averaged
annihilation rates. Such a calculation has apparently never been carried out and
is beyond the scope of this paper. Under the assumption of circular orbits, one
finds that the maximum number density at a given time $t$ can be expressed as
\be
\displaystyle
n_{sp}(r,t) \,=\, { n_{sp}(r,t_f) \over 1 \,+\, n_{sp}(r,t_f) \, (t - t_f) \,
\langle \sigma v \rangle } \,.
\label{rholim}
\ee
This is usually simplified to obtain a maximum density
\be
\displaystyle
\rho_{pl}(t) \,\approx\,  {m_\chi \over \langle \sigma v \rangle } {1 \over
(t-t_f)}
\ee
The radius where $\rho_{sp}$ reaches this value, denoted as $r_p$, can be
calculated by inserting Eq.~\refb{densitysp} into the above equation. The
maximum allowed density decreases with time due to self-annihilation and a
plateau of constant density forms from $r_p$ down to $r_{lso}$. As $r_{min}$ is
larger than $r_{lso}$, except for very massive halos, the  fully evolving spike
density profile is given as
\ba
\displaystyle
\rho_{sp}(r,t) \,=\, 
\begin{cases}
\rho_{pl}(t) &\, (r_{lso} \,<\, r \,\leq\, r_p)\\
\rho_{sp,0} \, \kappa^\epsilon \, \left({r \over r_b} \right)^{-\gamma_{sp}} 
&\, (r_p \,<\, r)
\end{cases} .
\label{den-sp1}
\ea

\begin{figure}
\begin{center}
\epsfig{file=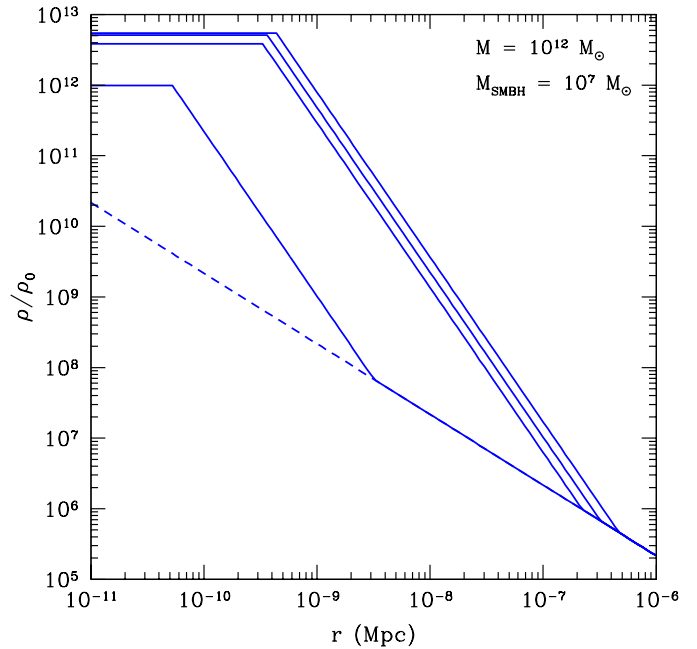,width=0.5\textwidth}
\end{center}
\vskip -2em
\caption{Density profile of an evolving spike. Halo and SMBH mass are fixed at
$10^{12} \,M_\odot$ and $10^7 \,M_\odot$, respectively, at all redshifts, and 
$z_{BH} = 6$. The dashed line is the halo profile, and solid lines are the spike
profile in various stages of evolution. Spikes are plotted at $\tau = \,
1,~2,~3,$ and $\approx 14 \,(z=0)$, from right to left (in order of decreasing
size). The spike becomes prominent at $r_b(t)$, which decreases with time due to
DM-baryon scattering. Self-annihilation of DM causes the maximum spike density
value of $\rho_{pl}$ to decrease with time. A constant density of DM with value
$\rho_{pl}$ is maintained from $r_p$ to the last stable orbit ($\sim 10^{-12}$
Mpc).}
\label{figspike}
\end{figure}
Figure \ref{figspike} shows the density profile of an evolving spike which has
formed at $z_{BH} = 6$. Note that the evolution of the halo itself has not been
taken into account and the halo and SMBH mass are fixed to $10^{12} \,M_\odot$
and $10^7 \,M_\odot$, respectively, at all redshifts in order to show only the
changes due to the evolving spike. The halo profile is plotted in dashed line
and the spike profiles at various $\tau$s are plotted in solid lines. The DM
profile is divided into three regions; a plateau with magnitude $\rho_{pl}$ from
$r_{lso}$ to $r_p$, the prominent spike that scales as $r \sim r^{-\gamma_{sp}}$
from $r_p$ to $r_b$, and the prominent halo with $r \sim r^{-\gamma_c}$ from
$r_b$ to $r_v$. Numerical computations, such as Ref.~\cite{Bertone:2005hw} which
has been calculated for our Galaxy, show the same features but with a smoother
transition at $r_b$.

It is convenient to express $\rho_{sp,0}$ in terms of $r_0$ and $\rho_0$, thus
of halo mass. Given the halo mass $M$, one can obtain $r_v$ from 
Eq.~\refb{virial} and $r_0$ from the definition of $c$ of Eq.~\refb{concen}. It
should be noted that $r_0$ is not a constant but varies with $z$ and $M$.
Similarly, $\rho_0$, which is also dependent on $z$ and $M$, is obtained by
solving
\be 
M \,=\, 4 \, \pi \int^{r_v}_0 dr \, r^2 \rho_h(r) \,.
\ee 
The reference spike density $\rho_{sp,0}$ is normalized by the halo density at
$r = r_{b,0}$ and $\tau = 0$. For a NFW halo this gives
\be
\displaystyle
\rho_{sp,0} \,=\, {\rho_0 \over x_{b,0} \, (1 \,+\, x_{b,0})^2} \,,
\ee
where $x_{b,0} \equiv r_{b,0}/r_0$, thus
\be
\displaystyle
\rho_{sp}(x) \,=\, \rho_0 \, {x_{b,0}^{\gamma_{sp}-1} \over (1+x_{b,0})^2 } \,
\kappa \,  x^{-\gamma_{sp}} \,. 
\label{eq:rhospx}
\ee

The expression of the total density profile depends on the redshift and radius.
The total density profile is given as follows:
\ba
\rho_{tot} \,=\, 
\begin{cases}
\rho_h (r) &  (z > z_{BH}) \\
\rho_h (r) & (z \leq z_{BH}, ~r > r_b) \\
\rho_h(r) \,+\, \rho_{sp}(r,t) & \\
~~~~\approx \rho_{sp}(r,t) & (z \leq z_{BH}, ~r \leq r_b)
\end{cases} \,.
\ea

\section{Gamma-Ray background from DM annihilations}
\label{sec:bgnd}

The contribution of DM annihilations to the gamma-ray background flux $\Phi$ 
can be expressed as ~\cite{Bergstrom:2001jj}
\begin{multline}
\displaystyle
\Phi(E) \,=\,  {c \over 4 \, \pi} {1 \over H_0}  { \langle \sigma v \rangle
\over 2} \left( {\Omega_m \, \rho_c \over m_\chi} \right)^2\, \times \\
\int dz \, {(1+z)^3 \over h(z) }  {dN(E_s) \over dE_s } e^{-\tau(z,E)} \zeta(z)
\,,
\label{flux}
\end{multline}
where $c$ here is the speed of light, $H_0$ is the present value of the Hubble
parameter, $E_s = E (1+z)$ is the energy emitted at the source, and $h(z) =
\sqrt{(1+z)^3 \Omega_m \,+\, \Omega_\Lambda}$. To allow an easy comparison with
existing literature, we adopt a simple analytic fit to  the continuum gamma-ray
flux emitted per annihilation coming from hadronization and $\pi^0$ decay
\cite{Bergstrom:2001jj}
\be
\displaystyle
{dN(E) \over dE} \,\approx\, {0.42 \over m_\chi} \, {e^{-8{E/m_\chi}} \over
(E/m_\chi)^{1.5} \,+\, 0.00014} \,,
\label{dnde}
\ee 
which is valid for $E \leq m_\chi$. The exponential in the integrand takes into
account the effect of gamma-ray absorption due to pair production on
background photons. Following \cite{Bergstrom:2001jj}, we write it as
\be
\displaystyle
e^{-\tau(z,E)} \,\approx\, 
\exp \left[ \frac{-z}{3.3 \, (E / 10 \, {\textrm{GeV}})^{-0.8}} \right] \,.
\ee

The dimensionless {\it flux multiplier} $\zeta(z)$ can be written as the
integral over all masses of an auxiliary function $g(M,z)$, weighted by the halo
mass function $dn/dM$, which is typically calculated in the framework of the
Press-Schechter or Sheth-Tormen formalisms described in 
Sec.~\ref{sec:halomassdistri},
\be
\displaystyle
\zeta(z) \,=\, \int dM \, {dn \over dM} \, g(M,z) \,.
\label{eq:zeta}
\ee
The auxiliary function $g(M,z)$ is simply the flux multiplier relative to a halo
of mass $M$ at redshift $z$, 
\be
\displaystyle
g(M,z) \,=\, {1 \over (\rho_c \Omega_m)^2 } \int_V dV \, \rho^2(r) \,,
\label{eq:gmz}
\ee
normalized to the comoving background density squared. $V$ is the halo virial
volume, which is a function of redshift, and of the halo mass and concentration
(see Eq.~\refb{virial}).

The integration over DM spikes requires particular care. Since we are assuming
that SMBHs do not evolve after their formation redshift $z_{BH}$, the halo
parameters in the $\zeta(z)$ calculation have to be evaluated at $z_{BH}$, while
the spike evolves with redshift as discussed above. Furthemore, the $M-M_{SMBH}$
relationship must evidently break down at small masses. Here we have restricted
the anlysis to spikes produced by BHs with mass $M_{SMBH} \geq 100 \, M_\odot$,
and have verified that the result is insensitive to this lower mass cutoff.

\begin{figure}
\begin{center}
\epsfig{file=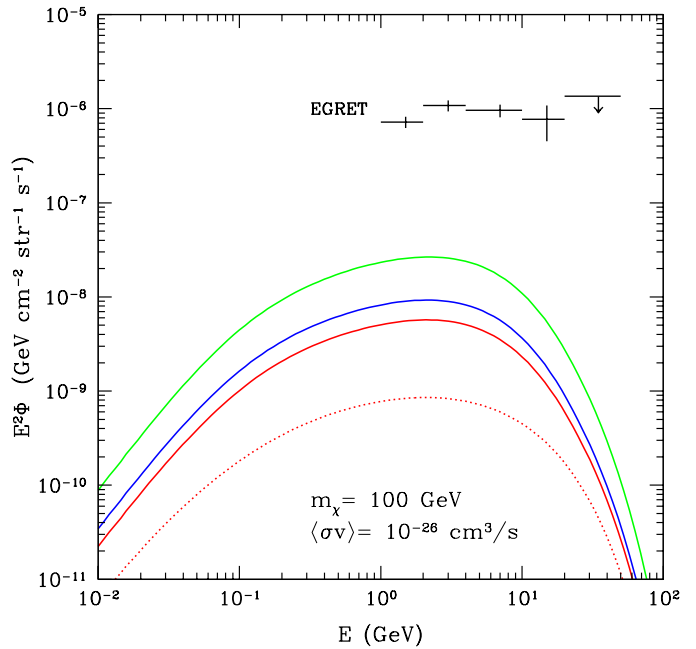,width=0.5\textwidth}
\end{center}
\vskip -2em
\caption{Gamma-ray background produced by DM annihilations in DM halos with
spikes (solid lines), compared to the halo-only contribution (dotted). The
EGRET diffuse flux limits \cite{Strong:2004ry} are shown for comparison. Halos
have mass between $10^5 - 10^{14} \,M_\odot$ and are distributed according to
the Press-Schechter formalism. The three expressions in Eq.~\refb{msmbhm} are
used for the $M_{SMBH}$-$M$ relation: $(a)$ (red) $(b)$ (blue) $(c)$ (green)
from the bottom to top. The DM parameters adopted are $m_\chi = 100$ GeV,
$\langle \sigma v \rangle = 10^{-26} \textrm{cm}^3 \textrm{s}^{-1}$.} 
\label{figflux}
\end{figure}
Figure \ref{figflux} shows the enhancement of the gamma-ray background due to
the presence of spikes, compared with the standard calculation (halo only). For
the figure, we have focused on the Press-Schechter formalism, but we find
similar results for the case of ellipsoidal collapse  {\it \`{a} la} Sheth \&
Tormen. We given an upper limit of $z=18$ to Eq.~\refb{flux} and assume that
spikes form at $z_{BH} = 2$. DM parameters $m_\chi = 100$ GeV, $\langle \sigma
v \rangle = 10^{-26} ~ \textrm{cm}^3 \textrm{s}^{-1}$ are used in the
calculation. All three mass relations between the SMBH and halo of
Eq.~\refb{msmbhm} are shown in the figure, $(a)$ to $(c)$ from bottom to top in
solid lines. The diffuse EGRET flux \cite{Strong:2004ry} is plotted as a
comparison. Enhancement due to the presence of evolving spikes is about
$\gtrsim$ order of magnitude. The evolving spike gives the largest enhancement
to the overall flux at the lower energy region, while there is little
enhancement at high energies close to $m_\chi$. This is expected, as the spikes
are most prominent just after formation at $z_{BH}$, and gamma-rays emitted
then have been redshifted to lower energies. Except for massive halos, most
spikes today have died away and contribute very little to the gamma-ray flux.
This also implies that the annihilation signal from the GC is not expected to
vary significantly from the case of profiles without spikes.

We also consider a case where gamma-rays are emitted from annihilation of
neutralinos into two photons. The photon flux is described as a delta function
where $dN(E)/dE = b_{\gamma \gamma} \delta(E - m_\chi)$, with $b_{\gamma
\gamma} = 0.003$ \cite{Bergstrom:2001jj}. Considering only a delta function as
the flux source is helpful to understand the enhancement due to the presence of
spikes.
\begin{figure}
\begin{center}
\epsfig{file=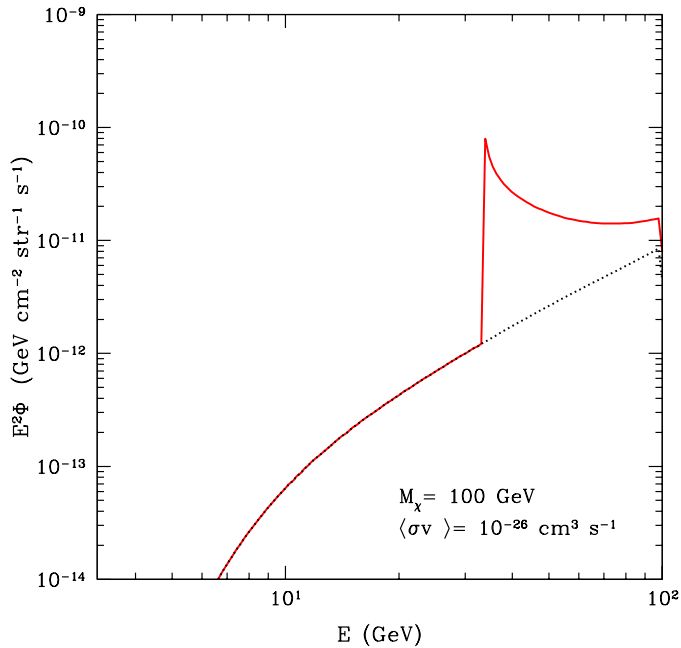,width=0.5\textwidth}
\end{center}
\vskip -2em
\caption{Gamma-ray flux due to annihilation of neutralinos into two photons.
Halos have mass between $10^5 - 10^{14} \,M_\odot$ and are distributed
according to the Press-Schechter formalism. Flux from halo only are shown in
black dotted line, and flux from halo and evolving spike are shown in 
red solid line. Spikes give the greatest contribution at low energies, 
i.e. at high redshifts.}
\label{fig:delta}
\end{figure}
The spike is expected to give the largest contribution around $z_{BH}$, which
today will be observed as gamma-rays with energy lower than $m_\chi$. Figure
\ref{fig:delta} shows the flux from annihilation into two photons for halo only
and halo and spike contributions, and indeed, the largest enhancement comes at
low energies. We have again assumed that halos form at $z=18$ and spikes form at
$z_{BH}=2$ and used the $M-M_{SMBH}$ relation Eq.~\refb{msmbhm}-(a). The steep
enhancement for the spike's flux at $E \approx 30$ GeV is due to our assumption
of having a fixed SMBH formation epoch ($z_{BH})$ and only using the delta
function for the gamma ray flux.

\begin{figure*}
\begin{center}
\epsfig{file=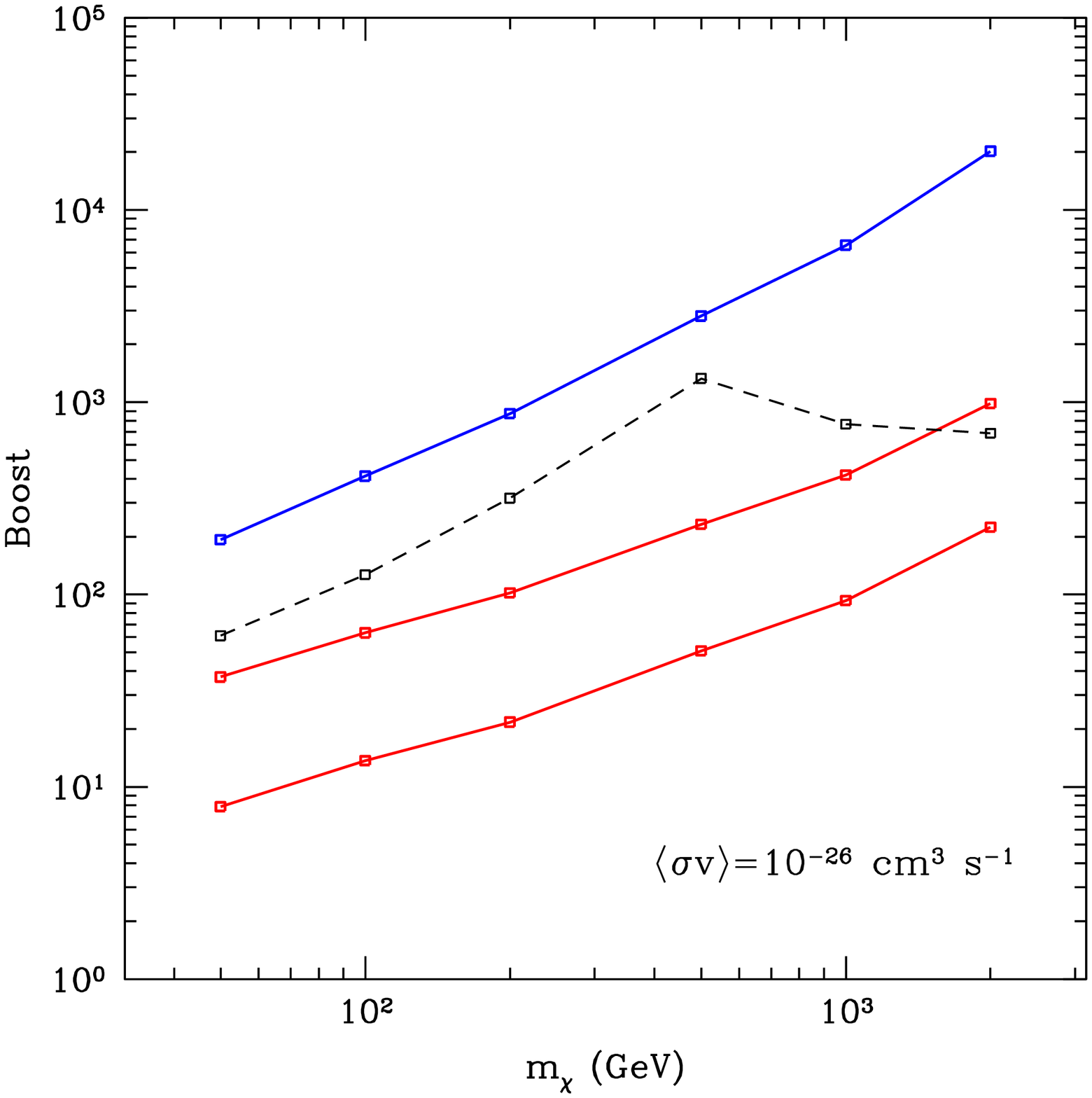,width=0.48\textwidth}
\epsfig{file=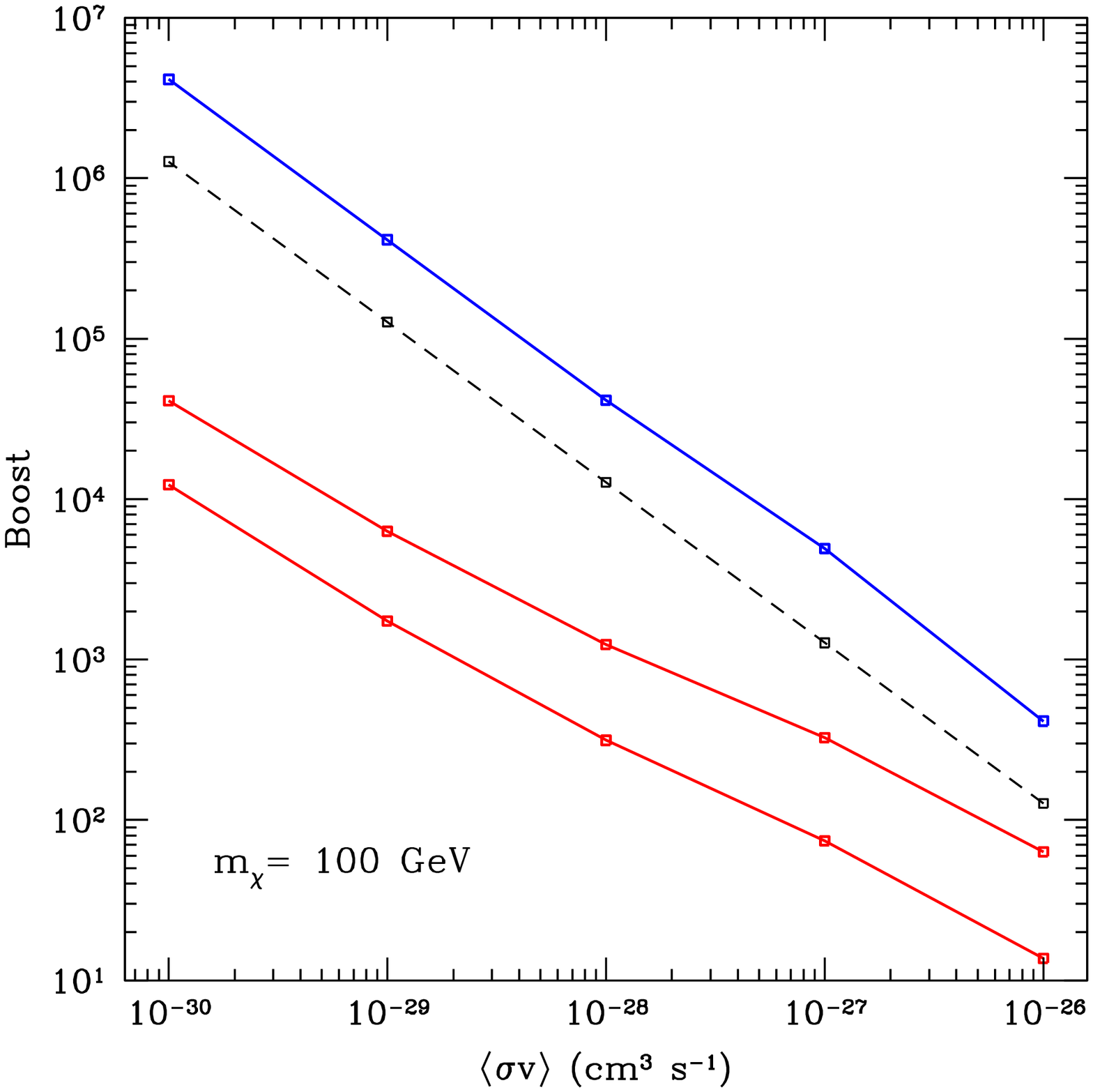,width=0.48\textwidth}
\end{center}
\vskip -2em
\caption{Required boost factor to match the EGRET background measurements with
DM annihilations, for the halo only case (top blue solid line); halo+spike with
$M_{SMBH}-M$ relation (a) (middle red solid line); and halo+spike  with
$M_{SMBH}-M$ relation (c) (lower red solid line). Halos have mass between $10^5
- 10^{14} \,M_\odot$ and are distributed according to the Press-Schechter
formalism. Left panel: $m_\chi$ varies while the cross section is fixed at
$\langle \sigma v \rangle = 10^{-26} \textrm{cm}^3 \textrm{s}^{-1}$. Right
panel: $\langle \sigma v \rangle$ varies with a constant $m_\chi = 100$ GeV. For
comparison, we show the boost factor relative to the gamma-ray source at the GC
(dashed line), using a NFW profile. Both EGRET and HESS observations are used.}
\label{figboost}
\end{figure*}
To compare with existing literature, we introduce here a ``boost factor'' 
$b_{max}$ defined as in Ref.~\cite{Ando:2005hr}, i.e. $b_{\rm max} \equiv 
\min_i [\Phi_{\rm EGRET}(E_i)/\Phi(E_i)]$, where $\Phi_{\rm EGRET}$ is the
EGRET flux measurement (Ref.~\cite{Strong:2004ry} for the diffuse background,
and Refs.~\cite{Mayer-Hasselwander98, Hartman:1999fc} for the GC) relative to
the energy bin $E_i$. We show in Fig.~\ref{figboost} the required boost factor
for 3 different cases: halo only (top solid line), and halo+spike with 2
different assumptions for the $M_{SMBH}$-$M$ relation (lower solid lines). The
left panel shows a constant $\langle \sigma v \rangle = 10^{-26} \textrm{cm}^3
\textrm{s}^{-1}$ with varying $m_\chi$ and the right panel shows a constant
$m_\chi$ with varying $\langle \sigma v \rangle$. The boost factor for the GC
is also shown as a comparision in dashed lines, where the HESS GC observation
\cite{Aharonian:2004wa} has also been considered. As one can see, for the
$M_{SMBH}$-$M$ relation of Eq.~\refb{msmbhm}-(c), the required boost factor for
the gamma-ray background is smaller than for the GC for most cases. We recall
here that the spike contribution scales differently with the particle physics
parameters $m_\chi$ and $\langle \sigma v \rangle$ with respect to the halo
only case, due to the saturation effects produced by annihilation itself. In
order for annihilations to contribute significantly to the observed gamma-ray
background, a boost factor of at least 2 orders of magnitude is thus required.
This could in principle be achieved by steepening the halo slope in the
innermost regions, for e.g. due to adiabatic compression of baryons (see e.g.
Ref.~\cite{Bertone:2005xv} and references therein), or to the presence of
mini-spikes around intermediate mass black holes \cite{Bertone:2006nq,
Bertone:2005xz}. One should however bear in mind that astrophysical sources are
expected to provide a significant, possibly dominant, contribution to the
background. Furthermore, the estimate of the background measured by EGRET has
actually been recently questioned by several authors. We discuss in the next
section the uncertainties on the EGRET  measurements and on the possible
astrophysical intepretation, and in light of these uncertainties, we do not
attempt to fit the background with a combination of particle physics and halo
models, and limit ourselves to point out the importance of the role played  by
spikes in the estimates of the DM annihilation contribution to the
extra-galactic flux.

\begin{figure*}
\begin{center}
\epsfig{file=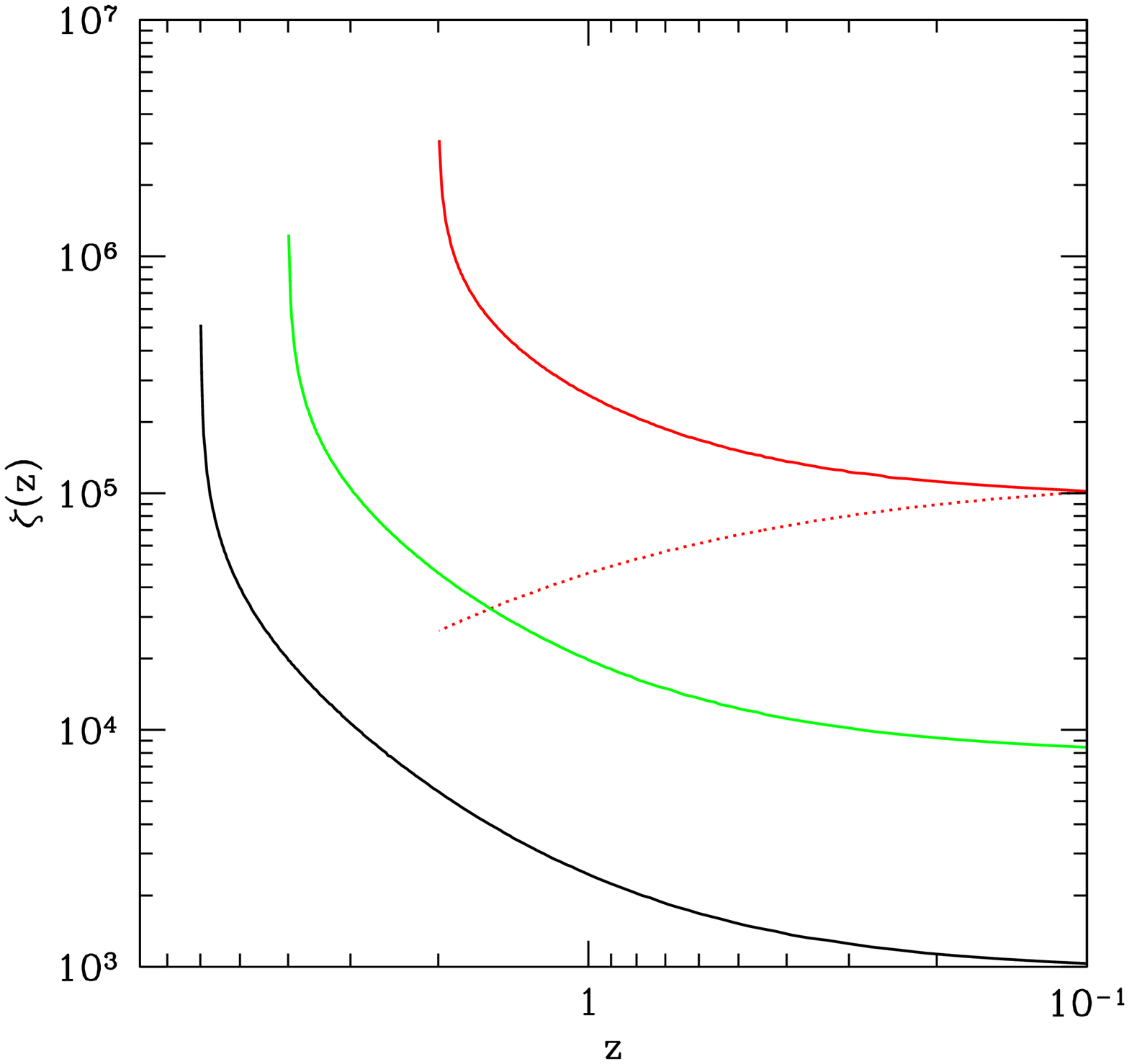,width=0.48\textwidth}
\epsfig{file=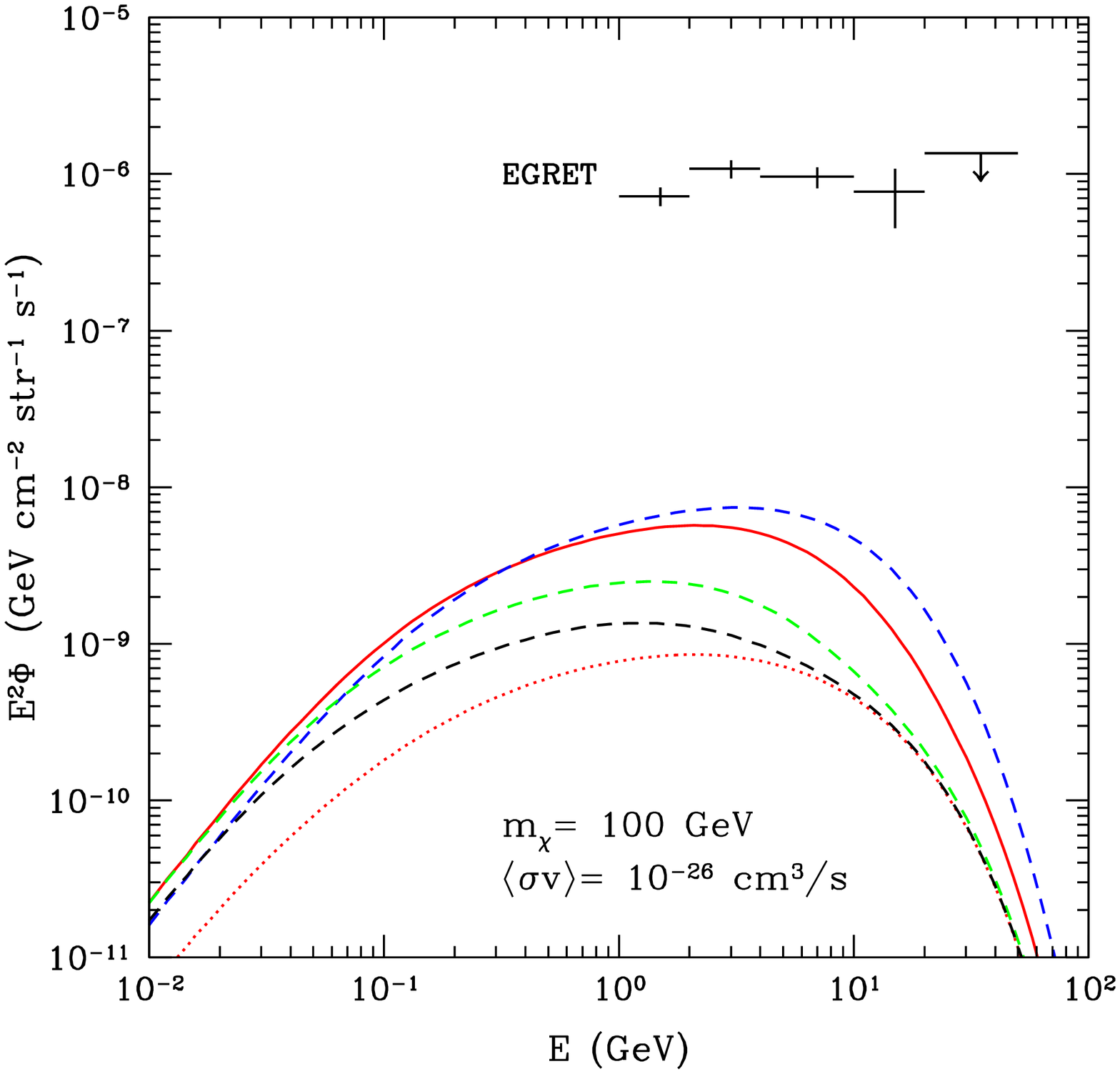,width=0.48\textwidth}
\end{center}
\vskip -2em
\caption{Left panel: $\zeta(z)$ for spikes (solid lines), assuming from bottom
to top, $z_{BH}=$ 6 (black), 4 (green), 2 (red). The halo $\zeta_z$
corresponding to the halo-only contribution is shown for comparion (dotted
line). Right panel: Sensitivity of the predicted gamma-ray background to the
formation redshift of SMBHs, $z_{BH}$, compared with $z_{BH}= 2$ case used in
previous figures. The lower dotted line represents the contribution of the halo
only, the four upper curves dashed and solid lines indicate the halo+spike
contribution for the $M_{SMBH}$-$M$ relation $(a)$ with, $z_{BH}=$ 6 (black), 4
(green), 2 (red solid), 1 (blue) from bottom to top. Halos have mass between
$10^5 - 10^{14} \,M_\odot$ and are distributed according to the Press-Schechter
formalism.} 
\label{fig:zbh}
\end{figure*}

In Figs.~\ref{figflux}-\ref{figboost} we have assumed a common redshift of
formation for all SMBHs. We show in Fig.~\ref{fig:zbh} the dependence of the
gamma-ray background on the parameter $z_{BH}$: the left panel shows the
evolution of $\zeta(z)$ (Eqn.~\refb{eq:zeta}) for different values of $z_{BH}$,
and the right panel shows the gamma-ray background. Younger spikes give a
greater contribution to the gamma-ray background because of a larger $\zeta(z)$.
The normalization of the annihilation signal has a slight dependence on
$z_{BH}$, where small $z_{BH}$ values give larger contribution to the gamma-ray
flux. This is expected as spikes that formed in earlier epochs evolve away with
time. 

On the other hand, dependence on halo formation redshift is negligible;
changing the upper limit on redshift fro Eq.~\refb{flux} from 18 to 12 brings
negligible change for both halo and halo+spike gamma-ray flux. The flux has a
small dependence on the maximum halo mass: a $10^{13} \ M_\odot$ limit lower
the flux by $<$ 20 \%. Varying the minimum halo mass brings negligible change.

\begin{figure}
\begin{center}
\epsfig{file=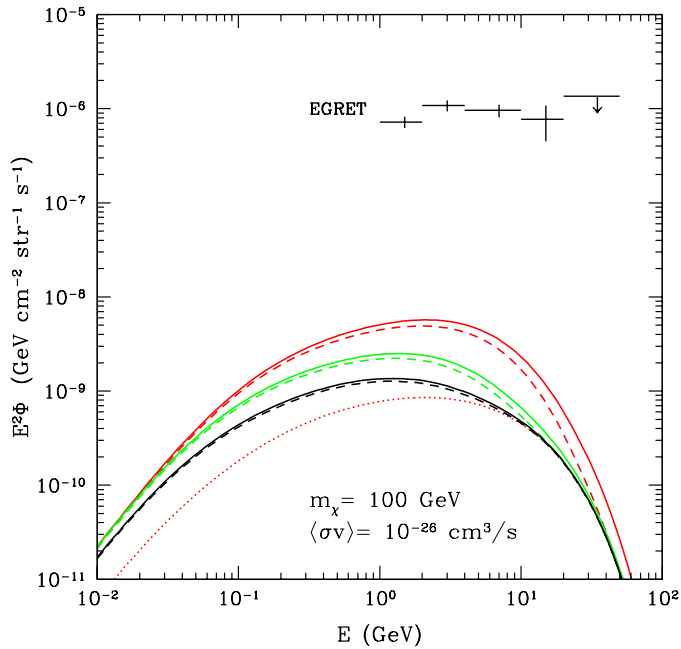,width=0.5\textwidth}
\end{center}
\vskip -2em
\caption{Gamma-ray background produced by DM annihilations in spikes including
(dashed lines) and neglecting (solid lines) the effect of mergers. Three
redshifts of formation have been considered for SMBHs; $z_{BH} =$ 2 (red, top
lines), 4 (green, middle lines), 6 (black, lower lines). The dotted line
represents the halo contribution. Halos have mass between $10^5 - 10^{14}
\,M_\odot$ and are distributed according to the Press-Schechter formalism.}
\label{fig:merger}
\end{figure}

All calculations so far assumed that spikes never experienced a major merger,
which could in principle significantly lower the DM density due to the souring
effect of binary BHs \cite{Merritt:2002vj}. To model the effect of galaxy
mergers on the  annihilation signal, we used the merger tree model of
Ref.~\cite{Wechsler:2001cs} (Eq.~\refb{halomassevol}), and assume that a galaxy
merger occured and its spike destroyed at $z_m$ when its halo mass at $z_{BH}$
doubles, i.e., $M(z_m) = 2 M(z_{BH})$. Figure \ref{fig:merger} shows the effect
merger has on the gamma-ray signals produced by spikes. The solid lines are the
spike+halo contribution without mergers, and the dashed lines are the
spike+halo contribution with merger taken into account. The dotted line is the
halo contribution only, shown for comparison. Three redshifts are considered,
$z_{BH}=$ 2, 4, 6, which gives a $z_m =$ 0.88, 2.12, 3.37, respectively.
The reason for such small effect of mergers can be seen from the left panel of
Fig~\ref{fig:zbh}; most of the contribution from the spikes come right after its
formation. By the time of the merger, most spikes would be already quite small
and not give significant contribution to the gamma-ray background.

\section{Discussion and conclusions}
\label{sec:conclu}

Different strategies have been proposed in the literature to search for DM
annihilation radiation. One of the most popular targets of indirect DM searches
is the GC. The prospects for detecting gamma-rays from DM annihilations at the
GC have been discussed extensively in Refs.~\cite{Bouquet:1989sr,
Stecker:1987dz, Berezinsky:1994wv, Bergstrom:1997fj, Cesarini:2003nr,
Bertone:2002ms, Cheng:2002ej} for DM cusps in the framework of different DM
candidates, and in  Refs.~\cite{Bertone:2002je,Bertone:2005hw,Bertone:2005xv}
for  the case of a DM spike at the GC. An updated discussion in light of the 
recent discovery of a point source coincident with the GC, extending to very
high energies can be found in Refs.~\cite{Hooper:2004vp, Profumo:2005xd,
Mambrini:2005vk, Zaharijas:2006qb}. Current data do not allow a convincing
interpretation of the gamma-ray  emission as due to DM annihilation, while the
properties of the gamma-ray emission appear consistent with those expected for
an ordinary astrophysical  source. However, although the DM intepretation of
the gamma-ray  source at the GC appears problematic, it can certainly be used 
in a conservative way to set upper limits on the annihilation signal. 
Alternatively, one could search for the contribution of DM annihilations to
the  cosmological gamma-ray background, as discussed in 
Refs.~\cite{Bergstrom:2001jj, Taylor:2002zd, Ullio:2002pj, Elsaesser:2004ck,
Elsaesser:2004ap, Ando:2005hr, Oda:2005nv}. The ``smoking-gun'' in this case
may come from the peculiar angular power spectrum predicted for this
signal~\cite{Ando:2005xg}. 

The predicted gamma-ray background is usually compared with the extra-galactic
emission measured by EGRET \cite{Sreekumar:1997un}. The most convincing
interpretation in terms of conventional astrophysical sources invokes a large
contribution from unresolved blazars (e.g. Ref.~\cite{Stecker:1996ma}), although
this conclusion has been challenged by other authors (e.g.
Refs.~\cite{Mukherjee:1999it, mucke}). An additional contribution may arise from
Inverse Compton scattering of electrons accelerated at shocks during structure
formation \cite{Colafrancesco:1998us, Totani:2000rg, Loeb:2000na}, but this
process can hardly account for the bulk of the background \cite{Gabici:2002fg}.
The EGRET extra-galactic background should however be treated with caution,
since it has been {\it inferred} (not {\it measured}) by substracting the
estimated Galactic foreground from the high latitude EGRET measurements. In a
recent re-analysis of the EGRET data, Kehet et al.~\cite{Keshet:2003xc}, noticed
that the high latitude profile of the gamma-ray data exhibits strong Galactic
features and claimed that it is well fit by a simple Galactic model, obtaining
an upper limit on the extra-galactic background 3 times stronger than previously
assumed, and evidence for a much lower flux. In light of the large uncertainties
associated with the data and with the contribution of conventional astrophysical
sources, we conservatively consider the EGRET estimate as an upper limit  to the
actual gamma-ray background, and do not attempt to fit the data with an {\it ad
hoc} combination of particle physics and halo properties.

A comparison of the two strategies (GC vs. extra-galactic background) has been
performed in Ref.~\cite{Ando:2005hr}, where it was shown that for ordinary
cusps, and in particular for an NFW profile, the prospects for detecting
gamma-rays from the GC are always more promising than for the gamma-ray
background. Here we have shown that the situation changes, if we take into
account the formation and evolution of DM spikes, which form due to adiabatic
growth of SMBHs at the centers of DM halos. In fact, in this picture a spike
inevitably develops also at the center of the Galaxy, but it is rapidly
destroyed by the combined effect of DM scattering off stars, and DM
annihilations themselves. The enhancement of the annihilation signal is thus
negligible \cite{Bertone:2005hw, Bertone:2005xv}. 

We have shown here that the opposite is true for the gamma-ray background. In
fact, although all spikes are affected by the very same processes, the signal in
this case receives contributions  also from halos at high redshift, at a time
when astrophysical and  particle physics processes did not yet have the time to
affect the DM  density. As a consequence, the gamma-ray background from DM
annihilations receives a substantial boost, so that its detectability is in some
scenarios more promising than the case of a gamma-ray source at the GC. An
additional reason to consider the gamma-ray background as a valid alternative to
GC searches, is that it is sensitive to the average properties of halos, whereas
in the GC case one has to deal with a single realization that, as far as we
know, may differ significantly from the average, given the significant scatter
in the properties of halos observed in numerical simulations, and given the 
unknown history of the baryons.

Several effects could further boost  the annihilation background. One
possibility is that DM halos undergo ``adiabatic contraction'' under the
influence of baryons, thus steepening the  original DM profile (see
e.g.~\cite{Bertone:2005xv} and references therein), a circumstance that would
lead to a similar boost of both the background and GC fluxes. Conversely, if
mini-spikes of DM around intermediate mass black holes 
exist~\cite{Bertone:2006nq, Bertone:2005xz}, this would have dramatic
implications for the predicted annihilation background, while leaving
practically unchanged the predictions for the GC \cite{Horiuchi:2006de}.

\section{Acknowledgements}

We thank S.~Ando, P.~Natarajan, L.~Pieri, G.~Sigl, R.~Somerville, M.~Volonteri,
and the anonymous referee for useful comments. The work of EJA was supported in
part by NSF PHY-0114422. KICP is a NSF Physics Frontier Center. The work of EJA
is now supported by the U.S. Department of Energy under Contract No.~DE-FG02
91ER 40626. The work of GB and PJZ was supported at an earlier stage of the
collaboration by the DOE and NASA grant NAG 5-10842 at Fermilab. GB is now
supported by the Helmholtz Association of National Research Centres. This
research was supported in part by the National Science Foundation under grants
no. PHY99-07949, AST-0206031, AST-0420920 and AST-0437519, by the National
Aeronautics and Space Administration under grant no. NNG04GJ48G, and by the
Space Telescope Science Institute under grant no. HST-AR-09519.01-A to DM.

\bibliographystyle{h-physrev3}
\bibliography{revise-dm-spikes}

\appendix
\section{Properties of DM Halos}
\label{sec:halo}

We present here our prescription to calculate DM density profiles $\rho_h(r)$,
for halos of mass $M$ and redshift $z$, and the dimensionless flux multiplier
$\zeta(z)$, that we used for the calculation of the gamma-ray background. For
the sake of completeness, we explicitly write all the ingredients of the
calculation, with relevant references, also to allow the comparison with
exisiting literature.

\subsection{Halo density profile}
\label{sec:hdp}

The virial mass $M$ of a DM halo can be expressed in terms of the virial radius
$r_v$ as
\be
\displaystyle
M \,=\, {4 \pi \over 3} \, r_v^3 \, \Delta_c \, \rho_c \,,
\label{virial}
\ee
where $\rho_c$ is the critical density and $\Delta_c$ is the virial overdensity,
that for a flat $\Lambda$CDM cosmology can be approximated by
\cite{Bryan:1997dn} 
\be
\Delta_c \,=\, 18\pi^2 \,+\, 82(\Omega_m(z) -1) \,-\, 39(\Omega_m(z) - 1)^2 \,.
\label{deltac}
\ee
Here, $\Omega_m(z)$ is the matter density in units of $\rho_c$.   

N-body simulations suggest that the density of DM follows a {\it universal}
profile, usually parametrized as~\cite{Bergstrom:2000pn, Bertone:2004pz}
\be
\rho_h(r) \,=\, \rho_0 \, { r_0^{\alpha \,+\, \beta \gamma} \over r^\alpha \,
(r^\beta \,+\, r_0^\beta)^\gamma } \,.
\label{densitypower}
\ee
where $\rho_0$ and $r_0$ set the normalization of the density and radius, 
respectively. We adopt here the so-called Navarro, Frenk and White (NFW)
profile \cite{Navarro:1996gj}, which can be obtained from the above
parametrization with the following choice of parameters $\alpha = \beta = 1,~
\gamma = 2$. In this case, the profile reduces to
\be
\rho_{NFW}(x) \,=\, {\rho_0 \over x\, (1 \,+\, x)^2}\,,
\label{densitynfw}
\ee
where $x \equiv r/r_0$. The reference density $\rho_0$ can be expressed in this
case as
\be 
\rho_0 \,=\, {1 \over 4 \pi} \, {M \over r_0^3} \, {1 \over \ln(1+c) - c/(1+c)}
\,,
\ee
where $c$ is an adimensional concentration parameter
\be
\displaystyle
c \,\equiv\, {r_v \over r_0} \,.
\label{concen}
\ee
Although subsequent studies have refined the description of DM halos with
respect to the NFW profile, and more recent parametrizations  of the innermost
regions of galactic halos appear better suited to capture the behavior in the
small-radii limit (see e.g.~\cite{Navarro:2003ew, Reed:2003hp}), we  present our
results  for the NFW profile, which has emerged over the years as a benchmark
model for DM annihilation studies. The scenario and prescriptions described
here, however, can easily be extended to any DM profile. 

A convenient expression for $c$ at redshift $z$ for halo mass $M(z)$ was
derived in Ref.~\cite{Bullock:1999he} from a statistical sample of
high-resolution N-body simulations, containing $\approx 5000$ halos in the
range $10^{11}-10^{14} \,M_\odot$:
\be
\displaystyle
c \,=\, {9 \over 1+z} \, \left( {M(z) \over M_\star(z=0) } \right)^{-0.13}\,.
\label{eq:concent}
\ee
The collapse mass $M_\star(z)$ defines the mass that collapses to form at halo
at redshift $z$. For $\Lambda$CDM cosmology, $M_\star(z=0) \simeq 1.5 \times
10^{13} \, h^{-1} \, M_\odot$. A detailed way of obtaining $M_\star$ is
described in Sec.~\ref{sec:haloevol}. Concentration parameters of halos less
massive than $\sim 10^{10} M_{sun}$ have not been robustly measured in N-body
simulations, due to the required mass and force resolution. For concreteness, we
assume Eq.~\refb{eq:concent} to apply for any halos. The validity of
extrapolating Eq.~\refb{eq:concent} to small halos should be tested against
future simulations.

\subsection{Halo mass distribution}
\label{sec:halomassdistri}

In this subsection, we discuss another key ingredient for the calculation  of
the cosmological annihilation flux. We need now to specify the mass function of
DM halos, i.e. the number of objects of given mass M. The mass function of
halos are expressed in a universal form \cite{Press:1973iz}
\be
\displaystyle
{dn \over dM} \,=\, {\rho_M \over M} \, {d\nu \over dM} \, f(\nu) \,,
\ee
where $\rho_M = \rho_c \Omega_M$ is the comoving matter (background) density and
a new variable $\nu$ is defined
\be
\displaystyle
\nu \,\equiv\, {1.686 \over D(z) \, \sigma(M)} \,.
\ee
The linear equivalent of the over density at collapse for spherical collapse
model is 1.686, $\sigma(M)$ is the rms density fluctuation in a sphere with
mass $M$, and $D(z)$ is the linear density growth rate.

In this way, $f(\nu)$ for spherical collapse is expressed as
\be
\displaystyle
f(\nu) \,=\, \sqrt{{2 \over \pi}} \, e^{-\nu^2 /2} \,.
\label{ps}
\ee
A better fit to number density of halos can be obtained with the 
ellipsoidal collapse model derived by Sheth and Tormen \cite{Sheth:1999su}; 
\begin{equation}
\displaystyle
f(\nu) \,=\, A \sqrt{a}\, \left(1 \,+\, {1 \over {\nu^\prime}^{2q}} \right) \,
\sqrt{{2 \over \pi}} \, e^{{-\nu^\prime}^2 /2} \,,
\label{st}
\ee
where $\nu^\prime = \sqrt{a} \nu$. Numerical fits to simulations give $a =
0.707, ~ p = 0.3$ \cite{Jenkins:1997en}, and $A$ is obtained by normalising
$f(\nu)$.

$\sigma(M)$ is related to the power spectrum $P(k)$ by
\be
\sigma(M)^2 \,\propto\, \int d^3k \, \tilde{W}(kR)^2 \, P(k) \,,
\label{sigmam}
\ee
where $k$ is the wave number and $\tilde{W}(kR)$ is the filter function with
$R$ being the radius enclosing mass $M$. We choose a top-hat function for
$\tilde{W}(kR)$;
\be
\displaystyle
\tilde{W}(y) \,=\, {3 \over y^3} \, (\sin\, y \,-\, y\, \cos\,y) \,.
\label{tophat}
\ee
We adopt here a power spectrum
\be
P(k) \, \propto \, k^n \, T(k)^2 \,,
\label{powerspec}
\ee
where the Bardeen-Bond-Kaiser-Szalay (BBKS) transfer function
\cite{Bardeen:1985tr} is used for $T(k)$:
\begin{multline}
\displaystyle
T(k) \,=\, {\ln (1 + 2.34q) \over 2.34q} \, [ 1 \,+\, 3.89q \,+\, \\
(16.1q)^2  + \, (5.46q)^3 \,+\, (6.71q)^4 ]^{-1/4} \,,
\label{transfer}
\end{multline}
with $q \equiv k / (\Omega_m h^2)$ (Mpc). $P(k)$ and $\sigma(M)$ are normalized
by simulating $\sigma(M)$ in spheres of $R = 8/h$ Mpc, commonly known as
$\sigma_8$. We use $\sigma_8 = 0.9$ from the concordance model.

Finally, we provide an expression for the linear growth rate $D(z)$
\cite{Carroll:1991mt}, which can be expressed as 
\be
\displaystyle
D(z) \,=\, {1 \over 1 + z} \, {g(z) \over g(0)} \,,
\label{dz}
\ee
where
\begin{multline}
g(z) \,=\, {5 \over 2} \, \Omega_m(z) \, \times
\left[ \Omega_m(z)^{4/7} \,-\, \right. \\
\left. \Omega_\Lambda \,+ \left(1 + {\Omega_m(z) \over 2}\right)\,\left(1 +
{\Omega_\Lambda \over 70} \right) \right]^{-1} \,.
\label{dz-s}
\end{multline}

\subsection{Evolution of halos}
\label{sec:haloevol}

In order to assign the appropriate BH mass to a host halo, we need to evolve
back in time the halo mass, and calculate its mass at the redshift of formation
of the SMBH. We follow the
semi-analytic study of halo evolution with merger trees, carried out in
Ref.~\cite{Wechsler:2001cs}, and express the halo mass at $z_0$ as a function
of an earlier redshift $z_1$
\be
\displaystyle
M(z_0) \,=\, M(z_1) \, \exp \left[ {S \over 1 + z_c} \, \left( {1 + z_1 \over 1
+ z_0} -1 \right)  \right] \,.
\label{halomassevol}
\ee
The free paramter $S$ is proportional to the logarithmic slope of accretion
rate, $d log M /da$ where $a = 1/(1+z)$, and is usually set to 2. The collapse
redshift $z_c$ is defined according to Ref.~\cite{Bullock:1999he}, where the
collapse mass $M_\star$ is a fixed fraction of the halo mass $M$
\be
M_\star(z_c) \,=\, f \, M.
\ee
In $\Lambda$CDM cosmology, $f$ is typically 0.01. $M_\star(z)$ is defined such
that $\sigma(M_\star(z),z)=1.686$, or equivalently, $\nu=1$.  We solve the
above equation to find $z_c$ for each M.

By setting $\nu = 1$, we have
\be
D(z_c) \,=\, {1.686 \over \sigma(M_\star(z_c))} \,=\, {1.686 \over
\sigma(0.01M)} \,.
\label{sigmazc}
\ee
Hence $z_c$ is obtained by interpolating the expression $D(z_c)$ of
Eq.~\refb{dz}.

\end{document}